\documentclass[preprint,12pt,3p]{elsarticle}

\usepackage{amssymb}
\journal{Nuclear Physics B}

\begin{document}

\begin{frontmatter}

%\title{Sample article to present \texttt{elsarticle} class\tnoteref{label0}}
\title{Fast Fourier-Based Generation of the Compression Matrix for Deterministic Compressed Sensing }
%\tnotetext[label0]{This is only an example}

%\author[label1,label2]{First name \\ gh}
%\author[]{Name \\ St. Cloud State University \\ Department of Information Assurance \\ htakanet@stcloudstate.edu}
\author[]{Sai Charan Jajimi \\ Department of Information Assurance \\ St. Cloud State University \\ sjajimi@stcloudstate.edu}
%\address[label1]{Address One}
%\address[label2]{Address Two\fnref{label4}}

%\cortext[cor1]{I am corresponding author}
%\fntext[label3]{I also want to inform about\ldots}
%\fntext[label4]{Small city}

%\ead{author.one@mail.com}
%\ead[url]{author-one-homepage.com}

%\author[label5]{Author Two}
%\address[label5]{Some University}
%\ead{author.two@mail.com}

%\author[label1,label5]{Author Three}
%\ead{author.three@mail.com}

\begin{abstract}
The primary goal of this work is to review the importance of data compression and present a fast Fourier-based method for generating the deterministic compression matrix in the area of deterministic compressed sensing. The principle concepts of data compression such as general process of data compression, sparse signals, coherence matrix and Restricted Isometry Property (RIP) have been defined. We have introduced two methods of sparse data compression. The first method is formed by utilizing a stochastic matrix which is a common approach, and the second method is created by utilizing a deterministic matrix which is proposed more recently. The main goal of this work is to improve the execution time of the deterministic matrix generation. The execution time is related to the generation method of the deterministic matrix. Furthermore, we have implemented a software which makes it possible to compare different methods of reconstructing data compression. To make this comparison, it is necessary to draw and compare certain graphs, e.g. phase transition, the ratio of output signal to noise and input signal to noise, signal to noise output and also the ratio of percentage of accurate reconstructing and order of sparse signals for various reconstructing methods. To facilitate this process, the user would be able to draw his/her favorite graphs in GUI environment.
\end{abstract}


\end{frontmatter}

% \linenumbers

\section{Introduction}
\label{sec1}

One of the current global issues is information and data transfer and data storage. Data  compression for reducing memory space and lessening broad ban have become a significant problem \cite{burrows1994block, ziv1977universal, witten1987arithmetic, pennebaker1992jpeg, bell1990text, sayood2005introduction, salomon2004data}. Having effective compression algorithms, the storaed data on data centers can be alleviated, resulting in a faster inter-datacenter data communication at data centers \cite{noormohammadpour2017datacenter, yekkehkhany2017near, noormohammadpour2016dcroute}. Natural data resources such as people's speech and writing have extra characters that will not pronounce as well. For example, in the sentence ``I returned to our house'', the pronoun ``our'' and the identifier in the verb ``ed'' can be deleted in that sentence without reducing the conceptual meaning of the statement \cite{spolsky1969reduced, i2001small}. The same operation can happen in data compression. Therefore, data compression means reducing its volume in such a way that its content does not undergo improper changes. Similar to any communication, communicating with compressed data functions only when data transmitter and receiver understand the coding method. For example, this writing is merely meaningful when a receiver understands that the goal of implementation is achieved by using English language. Compressed data is useful when a receiver knows the right decoding method \cite{Wiki, berrou1993near, bahl1974optimal, hall2001encoding, viterbi1967error, gough1986decoding}.

In the area of data compression, we focus on sparse data which have vectors with a considerable number of zero \cite{daubechies2004iterative, candes2008enhancing, tibshirani2005sparsity}. In addition, other principle concepts are defined as follows. Stochastic and deterministic compression are discussed \cite{egiazarian2007compressed, rauhut2008compressed, applebaum2009chirp, devore2007deterministic}. Unique focus is on the production of deterministic matrices for sparse signal compression. The main goal of this work is to present a novel method to reduce the time for production of deterministic matrix by MATLAB software. Moreover, various reconstructing data compression methods are introduced and compared with each other. To carry out this comparison, one can compare the graphs related to these reconstructing methods such as graph of phase transition, the graph of the ratio of signal to output noise to input noise signal as well as the graph of ratio of signal to output noise to order of sparse signal $X_{n*1}$ and also the graph of ratio of the accurate reconstructing percentage to the order of sparse signal $X_{n*1}$. In chapter \ref{sec4}, we explain these graphs thoroughly. To draw these graphs, GUI environment is used to design without the need to know MATLAB software (M-file) method nor reconstructing methods.

\section{Principle Concepts}
\label{sec2}

In this chapter, we primarily define the principle concepts that are required for this work.

\subsection{General Method of Vector Compressing}
\label{subsec21}
As mentioned in the introduction, communication with compressed data is only possible when both sender and receiver of data understand the decoding method \cite{amiri2013data, ikelle2007coding, loia2005fuzzy}. In data compression, by utilizing the method under our study, we primarily multiply the matrix $\Phi_{m*n}$ (features will be explained further) by a sparse vector $X_{n*1}$ which we intend to compress in order to obtain a novel vector $Y_{m*1}$. If we assume $m<n$, the new and obtained signal $Y_{m*1}$ has a shorter length compared to the main signal. The above trend is illustrated in figure 1.

\begin{figure}
\center
\label{CS}
\includegraphics[scale=0.5]{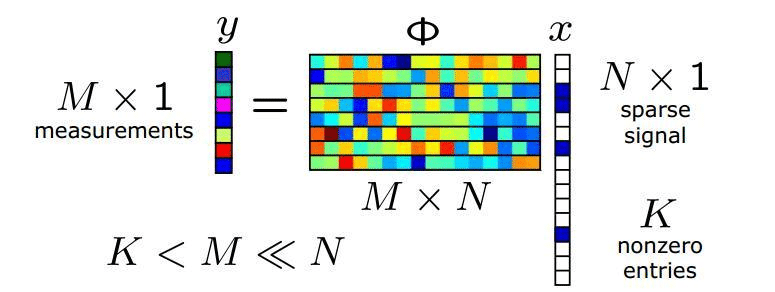}
\caption{The general picture of compression method of sparse data.}
\end{figure}

The operation of the main data retrieving is essential in data compression. If we can retrieve signal $X_{n*1}$ from compressed signal $Y_{m*1}$, a systematic method of data compression and retrieving is obtained for the data. As it will be explained further, without any condition over matrix $\Phi_{m*n}$ and vector $X_{n*1}$, retrieving the main data is not possible. In the following sections, we will discuss and describe this issue more precisely.

\subsection{Sparse Signals Definition}
\label{subsec22}

Sparse signal is commonly referred to a signal having numerous zero elements; in other words, its nonzero elements have a smooth and small signal. The physiology data studied in \cite{hosseini2017mobile} is a good example of sparse signals that are critical for transmission over communication channels. Failure in transmission of physiology sparse data can incur critical costs to patients \cite{hosseini2017mobile}. In this work, we use the data set provided by Hosseini et al. \cite{hosseini2017mobile} to test various deterministic compressed sensing methods on data provided by them, which can be used as a supplementary method to their route selesction method for speeding data transmission in ambulatory transportations. Next, we define sparse signal precisely. Matrix $X_{n*1}$ in the base of $\Psi$ is called $k$-Sparse if the relation is as given below \cite{baraniuk2007compressive, ba2010lower, shen2008sparse, zhang2011sparse}:

\begin{equation}
\label{equ21}
X_{n*1} =\Psi_{n*n} \alpha_{n*n} \ \ \ \textit{such that: } \| \alpha_{n*n} \|_0 \leq k.
\end{equation}

In fact, note that $\Psi$ should not have zero determinant in order to be used as a base operator \cite{Wiki2}.

In practical applications, usually $K << n$ in order to be able to ensure that the signal is sparse \cite{toint1977sparse, fletcher2009necessary, lorenz2008convergence, beck2013sparsity}.
An image of a sparse signal is illustrated in in figure 2.

\subsection{Definition of Matrix Coherence }
\label{subsec23}

In this section, we merely define matrix coherence terminology and point out its usage in the following subsections.
To calculate matrix coherence $\Phi_{m*n}$, it is necessary to obtain the maximum coherence between each two various columns. The coherence of two vectors is also defined as the absolute value of the total arrays of two similar vectors multiplied by each other \cite{pass1997comparing, wen2008image, hongli2004region, zhou2011measure}. We primarily calculate the two inner vectors by multiplying them and then derive its absolute value. If we perform this task for each two separate columns of matrix and find their maximum, matrix coherence is obtained \cite{drineas2012fast, talwalkar2014matrix, bjornerud1995flow, mahoney2012fast}.

\begin{figure}
\center
\label{Example}
\includegraphics[scale=0.5]{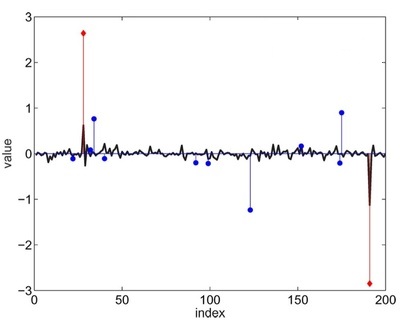}
\caption{An example of sparse signals.}
\end{figure}

\subsection{Restricted Isometry Property (RIP)}
\label{subsec24}
In this section, we also define another concept which is used further. Matrix $\Phi_{m*n}$ adhere to RIP from $k$ order having a constant value $0 \leq \delta_k \leq 1$ if we have $X_{n*1}$ $k$-Sparse for all vectors \cite{Wiki2, candes2008restricted, baraniuk2008simple, davenport2010analysis, krahmer2011new}:

\begin{equation}
\label{equ22}
1 - \delta_k \leq \frac{\| \Phi X \|^2_{l_2}}{\| X \|^2_{l_2}} \leq 1 + \delta_k.
\end{equation}

\subsection{Features of Matrix compressor $\Phi_{m*n}$}
\label{subsec25}
It is worth mentioning that there are two major problems in data compression. First, suitable selection of $\Phi_{m*n}$ for maximizing compression of sparse data, and the next problem is reconstructing the main data $X_{n*1}$ from compressed signal $Y_{m*1}$. Considering the general illustration of operation of compression in figure 1, assume the sparse signal $Y_{m*1}$ and matrix $\Phi_{m*n}$ are available and we intend to reconstruct the main signal $X_{n*1}$, a set of equations $m$ having unknown $n$, ought to be solved.
As mentioned previously, to have an acceptable compression, there ought to be $𝑚 << 𝑛$. However, in this case, the number of unknown is greater than the number of equations which in general, signal reconstructing will be impossible without setting any conditions over the problem. Therefore, we ought to set a condition for matrix $\Phi_{m*n}$ and sparse signal $X_{n*1}$ to create the possibility to reconstruct the data. With respect to the presented definition, if the signal which is intended to be compressed is sparse order $k$, and if matrix $\Phi_{m*n}$ adhere to RIP of order $2*k$ having suitable constant $\delta_{2k}$ as explained in previous section, it is possible to reconstruct signal $k$-Sparse from noise samples obtaining great results \cite{Wiki2}.

In addition, it can be shown that it is essential for matrix coherence to limit the signal sparse order $X_{n*1}$ to a certain limit (which the coherence depends on it). To generate this type of matrix columns, finite fields are used \cite{menezes1993reducing, reed1960polynomial, pollard1971fast, quillen1972cohomology}. In finite field mathematics, there is a field which is comprised of many finite elements. These finite fields are divided with respect to their size. For every measure of the field $pk$, a finite field is precisely definable. Note that, $p$ is the primary number and $k$ as a natural number. The finite fields are broadly applicable in many areas such as mathematics and computer science \cite{applebaum2009chirp}.
The MATLAB code written to produce deterministic matrix $\Phi_{m*n}$ is as follows. We used finite fields to produce matrix columns in such a way that fulfills the necessary conditions of output matrix and have low coherence and adhere to RIP from order of $2*k$ by suitable constant $\delta_{2k}$.

\section{Stochastic and Deterministic Compression of Data}
\label{sec3}
In previous chapter, we became familiar with necessary principle concepts. In this chapter, two general methods are presented to produce matrix $\Phi_{m*n}$. Our focus is on deterministic method. Since the stochastic method has extensive applications, we briefly explain it as well and point out some of its deficiencies.

\subsection{Production of Stochastic Matrix $\Phi_{m*n}$}
\label{subsec31}
As it was stated in chapter \ref{sec2}, matrix $\Phi_{m*n}$ supposed to be valid in a condition with respect to the degree of the sparse signal $X_{n*1}$. Traditionally, matrix $\Phi_{m*n}$ was generated stochastically \cite{donoho2006compressed, lustig2007sparse, candes2008introduction, candes2006compressive, tian2007compressed}. In fact, every matrix element used to be obtained by a Gaussian stochastic process. By using this method, a negligible value of matrix coherence is highly probable. However, there is no guarantee for the mentioned matrix to adhere to RIP by the order of $2*k$ with a suitable constant $\delta_{2k}$, and also have low coherence as well. Therefore, lack of certainty is observed in this method \cite{devore2007deterministic, applebaum2009chirp}. Furthermore, as stated in the introduction, a communication is possible by time compressed data only if both sender and receiver of data are aware of the details of compression and its method. During this process, it is necessary for both sender and receiver to know matrix $\Phi_{m*n}$ to reconstruct the data accurately \cite{li2012deterministic, calderbank2010construction, howard2008fast}. Since the sender generates a stochastic matrix during data compression, the receiver has no information about matrix and the sender supposed to transmit matrix $\Phi_{m*n}$ along with the compressed data to provide the possibility of reconstructing the main data. In every case mentioned above, the broad band or more memory is used to store $\Phi_{m*n}$, which is not desirable \cite{yu2013deterministic, li2011deterministic, bayati2011dynamics}. However, in the proposed method of generating the deterministic matrix, before data compression, matrix $\Phi_{m*n}$ is determined for a pair of sender and receiver and there is no need to send or store matrix $\Phi_{m*n}$. Note that the generation of the deterministic matrix $\Phi_{m*n}$ can be made fast by utilizing parallel processing of data and using sophisticated MapReduce algorithms, e.g \cite{dean2008mapreduce, dean2010mapreduce, chu2007map, yekkehkhany2018gb, zaharia2008improving, daghighi2017scheduling, ranger2007evaluating, yang2007map}.

\subsection{Generating deterministic matrix $\Phi_{m*n}$}
\label{subsec32}
As explained, by the method of generating deterministic matrix $\Phi_{m*n}$, finite fields are used to generate matrix $\Phi_{m*n}$ columns \cite{yu2013deterministic, li2011deterministic, xu2015compressed, li2014deterministic, candes2006near}. Note that not only the matrix coherence is smaller than the desired value but also Matrix $\Phi_{m*n}$ adhere to RIP from the $2*k$ order with suitable constant $\delta_{2k}$ as well. It is noteworthy that the matrix elements are no longer stochastic and are generated as systematic matrix $\Phi_{m*n}$ thoroughly.
At this point, the goal is not to explain how the matrix $\Phi_{m*n}$ is generated and for further information, see the article in the reference section \cite{Wiki2, candes2006robust}.
The main goal is to improve the MATLAB code which builds the matrix $\Phi_{m*n}$ and to increase the code speed as well as matrix $\Phi_{m*n}$ production. The old method of matrix $\Phi_{m*n}$ production takes a lot of time overhead in order to produce a large size of data. However, by the new method which will be introduced, this timing is reduced.

First, it is necessary to discuss about deterministic matrix $\Phi_{m*n}$. This matrix has an interesting feature as an assisting agent; if it is complete with respect to size, every matrix column is selected and shifted as circular and then a new circular and shifted vector can be found in one of the columns (circular shift of a vector means that the last vector element is taken and the remaining vector arrays are shifted toward bottom and the last old array is placed in the location of the first vector array). During matrix production by using finite fields, these shifted circular columns are not next to each other. They are usually distributed irregularly along the matrix. Furthermore, we will explain that if shifted circular columns are placed next to each other, we begin from the first matrix column and shift it circularly by a unit value. The outcome vector is in the second matrix column. Now if the second column is shifted circularly by a unit value, the next column which is the third column is generated. We continue this task until no new column is being generated. Therefore, the calculation volume is greatly reduced. Our desired goal is to create similar conditions for the remaining matrix columns. This means that if we select a desired matrix column and shift it circularly, it is precisely in the next column. In this case, the matrix is divided into blocks in which each block would be the shifted circular block of a vector. The general view of ordered and desired matrix is illustrated in figure 3.

\begin{figure}
\center
\label{ordered}
\includegraphics[scale=0.5]{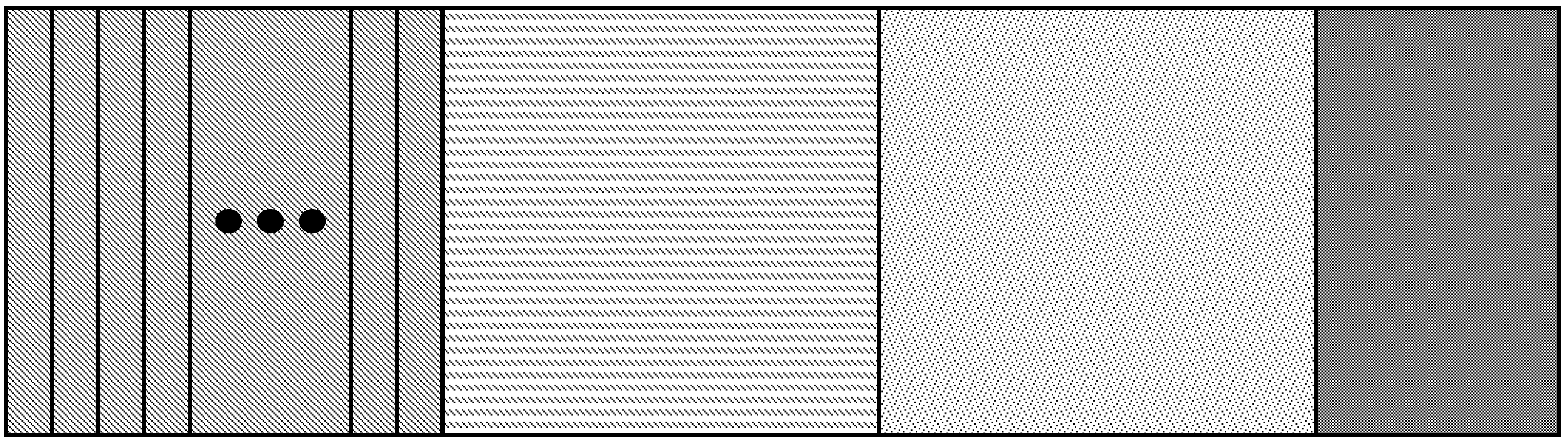}
\caption{General view of ordered matrix $\Phi_{m, n}$.}
\end{figure}

In this state, by using the properties of Fourier transfer, the calculation volume in reconstructing will be from $n\log(n)$ order. However, if matrix $\Phi_{m, n}$ is not ordered as blocks as explained, the calculation volume in reconstructing will be from $n^2$ order. The reason for having calculation volume reduction is that it is necessary to multiply compressed data $Y_{m*1}$ by transposes of matrix $\Phi_{m, n}$ in the process of reconstructing compressed data. In the case that we intend to carry on this multiplication ordinarily, the calculation volume is from $n^2$ order. So when the matrix $\Phi_{m, n}$ is separated in blocks as mentioned, we can remarkably reduce the calculation volume by using the properties of Fourier transform. Instead of multiplying every single column of matrix $\Phi_{m, n}$ by $Y_{m, 1}$, we obtain Fourier transform from the first column of every block and then multiply it by Fourier transform of signal $Y_{m, 1}$. The next step would be obtaining the opposite of Fourier transform from the result. As a result, the calculation volume is remarkably reduced. To obtain the Fourier transform, it is better to use the FFT command \cite{welch1967use, katoh2002mafft, carr1999option, van1992computational, brigham1988fast, cochran1967fast} instead of DFT command and for the opposite of Fourier transform, we utilized the IFFT command.

It came to our attention that sorting the matrix as mentioned, reduces the calculation volume remarkably. Therefore, we presently search for a method to sort the matrix quickly which would have a short execution time. In this work two methods are introduced. The first sorting method takes a great deal of time. The second sorting method is very fast and increases the speed of software considerably by utilizing Fourier transform.

\subsubsection{The first method of sorting matrix $\Phi_{m, n}$}
\label{subsubsec321}
As stated earlier, it is assumed that by using finite fields, the entire matrix $\Phi_{m, n}$ is available and we only intend to sort it into our desirable form. If matrix $\Phi_{m, n}$ is complete, the circular shift of the result of every column of matrix can be found in one of its columns. However, if the matrix is not complete, there is possibility that all the results circular shift is not found in the matrix.
Therefore, in the first method, the method starts from the first column of matrix and shift it circularly by a unit value, then it continues to the second column. Then the matrix should be searched for this column and if there is such column, it should be replaced. Moreover, if the mentioned column was not found, the first column is shifted twice circularly and the search process has to be done again and if the desired column was found, then place it in the third column of the matrix and delete it from the main matrix. Continue this task until the entire circular shift of the results of the first column are checked. Furthermore, following the same task for other columns until the entire matrix columns are sorted in their related blocks shows that a specific column and circular shift of results are orderly placed next to each other. However, this matrix sorting method is time consuming and has high execution timing. As it will be explained, by the method presented, the execution timing to sort the matrix $\Phi_{m, n}$ is highly reduced. Comparing the execution timing of both methods are presented in one graph.

\subsubsection{The second method of sorting matrix $\Phi_{m, n}$}
\label{subsubsec322}

As mentioned, we intend to sort matrix $\Phi_{m, n}$ in such a way that it is formed by blocks. Moreover, each one of circular shift of the result of vector belongs to one block. If we observe these blocks from frequency point of view, every block column has a common feature. We know that the measure of (abs) of Fourier transform vectors which are circular shift of the result and they are equal to one another. This means that identical vectors can be obtained by performing Fourier transform from the entire columns of the matrix $\Phi_{m, n}$, and each one of the Fourier transform results can be measured by using abs command. Therefore, we can use this final vector as the characteristic of each matrix block. In the next step, we intend to generate a matrix having a big data which its matrix coherence is lower than its data and adhere to RIP from $2*k$ order with suitable constant $\delta_{2k}$. They are sorted as blocks which are circularly shifted of separate resulted vectors.

In this method, it is not necessary to generate the entire matrix $\Phi_{m, n}$ at the beginning. We primarily generate a column of the matrix (This task is done by finite fields to find the precise method, see the references) and place it in the first column of the matrix $\Phi_{m, n}$. As we know, the first column entire circular shifts can be part of the matrix. Therefore, instead of receiving help from finite fields and its related software to generate the second column matrix $\Phi_{m, n}$ again, we shift the first column by a unit value manually and place it in the second column. Subsequently, we shift the second column by a unit value circularly and place it in the third column and continue this process until the first column has been achieved. In this stage, if the necessary length for matrix $\Phi_{m, n}$ was not generated, we generate a new column and place it next to other columns and shift it circularly as much as possible, and then set it next to one another. However, the issue is that there is a probability for the new column to be repetitive. In fact, when we generate the previous columns and shift them, it is possible for the new generated column to be among the columns that were circularly shifted. So we should check this problem that whether or not there is a new generated column among the entire generated columns alongside with circularly shifted ones. If we intend to check a new generated column one at a time, it is very time consuming. Especially considering the scenario in which the matrix size becomes large. Hence, the characteristic of every block has been used. As stated, the size of Fourier method of the entire columns belonging to a block are equal to one another. Therefore, after every production of a column it is recommended to assure that the column is not repetitive and shifted and would be placed next to each other. One vector is stored as representative of the mentioned block, which has the same norm as Fourier transform of one of the column of that block. Next, if we produce a new column, we perform Fourier transform on it and then its size would be calculated. The produced vector is compared with the characteristics of previous blocks produced (the characteristics of each block means the size of Fourier transform of one of the columns of that block). If the size of Fourier transform of the new column is equal to the characteristic of one of the previous blocks, we realize that the new and produced column is repetitive and we would have to generate a new column. However, if it is not equal to none of the blocks characteristics, we conclude that the new column produced is not repetitive and there is a possibility to add this column alongside with its circular shift to the matrix. By this method, the speed of production of matrix $\Phi_{m, n}$ is remarkably increased. In real time applications, one of the vital issues is to speed up the processing and data compression.
As shown in the graph below, the new method presented generates the desirable matrix at a very fast rate. In the graph below, the time it takes to produce the matrix based on various lengths is presented for both methods. It should be mentioned that in order to have the possibility to compare both methods appropriately, the matrix the same coherence $\Phi_{m, n}$ should be considered for both methods.
In figure 4, the time needed to produce the matrix by the first method is colored by blue and the time needed to produce the matrix by the new method presented is colored by red.
\begin{figure}
\center
\label{comparison}
\includegraphics[scale=0.25]{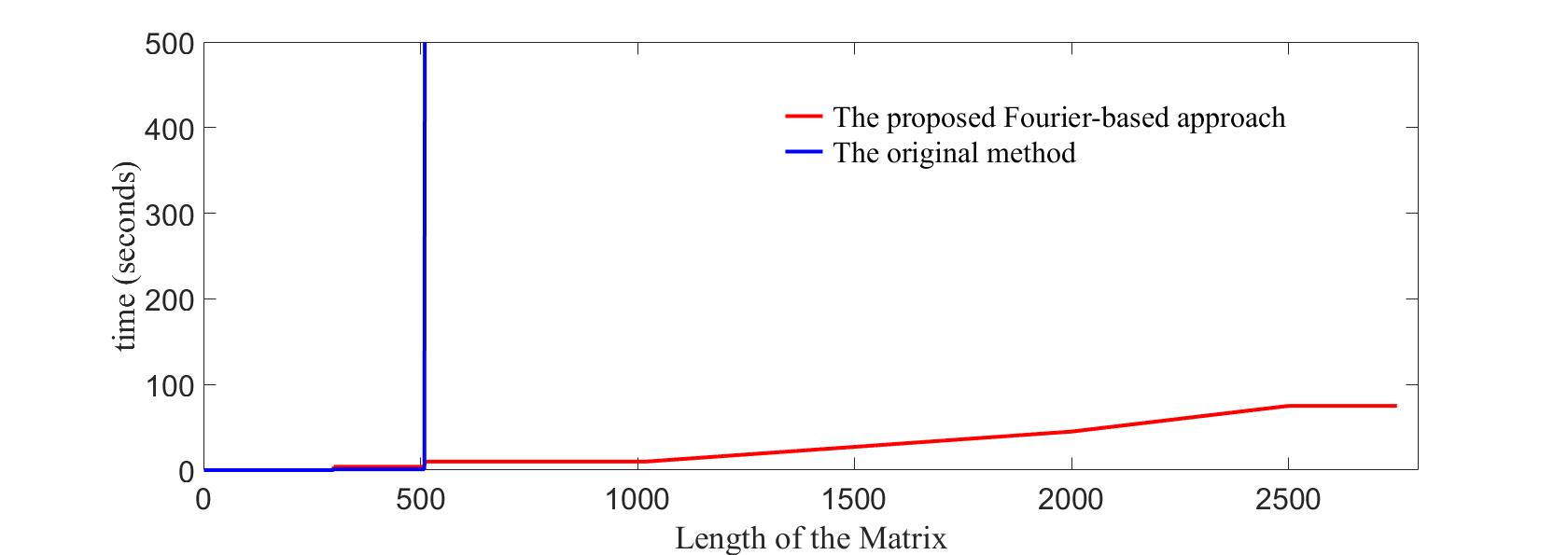}
\caption{Time comparison to execute the software between the previous method and the new method presented.}
\end{figure}
As it is observed, the time needed to generate the matrix by the new method is greatly smaller than the old method. Moreover, as the Matrix length grows the new method is showing better efficiency.

\section{Comparing Various Methods of Data Reconstructing}
\label{sec4}
To reconstruct the compressed data, there are various methods. To compare the efficiency of these methods, we have various criteria. Several common methods of data reconstructing were mentioned in \cite{Wiki5, zhang2009user}. Some of these methods are as follows to name but a few. OMP \cite{donoho2006stable}, CoSamp \cite{needell2009cosamp, needell2010cosamp, davenport2013signal, giryes2012rip, giryes2012cosamp, becker2012cosamp}, Yall1 \cite{yang2011alternating}, IHT \cite{blumensath2009iterative}, IMAT, OMP\_enhanced, AMP \cite{donoho2009message} and CoSamp\_enhanced. The last two methods are applicable when the matrix is generated deterministically and its columns are sorted as shown in figure 3. It is noteworthy that the last two methods have higher speed compared to other methods as they are used to compress and reconstruct data by deterministic matrix.

In the next section, we discuss the four criteria to compare the efficiency of aforementioned methods.

\subsection{Phase Transition}
\label{subsec41}
Assume $n$ to be a constant number for the length of compressor matrix $\Phi_{m*n}$. In this case, as stated previously, m shows the number of samples taken from vector $X_{n*1}$, which would be a vector from $k$-sparse order. It is evident that the larger number is $m$ and as a result, more samples are taken from vector $X_{n*1}$, the data reconstructing is done by a higher ratio of signal power to noise power. Sometimes we should make sacrifices in data compression with a higher ratio. Assuming no noise is collected by a compressed signal, it is expected that the ratio of signal power to noise to be infinitive. However, these reconstructing methods use algorithms which tend toward numerous or even infinitive repetitive answers. Consequently, there is no possibility to practically reconstruct data precisely \cite{wu2012optimal, dutta2018optimal}. On the other hand, MATLAB software has limited capability to store numbers that have many decimals, which is a factor to avoid obtaining the ratio of signal to high noises for reconstructed signals. Moreover, the lesser is the degree of sparseness of vector $X_{n*1}$ the easier it would be to compress the data in $X_{n*1}$ by a higher ratio which means by choosing a smaller $m$ for matrix $\Phi_{m*n}$ we will avoid losing the main signal data.

By the aforementioned introduction, we define the phase transition graph. Assume n to be constant number and the value of $m$ changes from $1$ to $n$. Then, we change the degree of sparseness of signal $X_{n*1}$ from $1$ to $m$. For every value of $m$ and $k$ $(k< m< n)$ the sparse vectors from $k$ order by numerous repetition are generated and we compress them by stochastic matrix $\Phi_{m*n}$. Subsequently, by using one of the reconstructing methods, the signal $Y_{m*1}$ would be reconstructed. Finally, with respect to the main signal being available, we obtain the ratio of signal power to noise power for reconstructed signal. If signal to noise is more than a certain value (threshold), we consider the reconstructing to be accurate. Otherwise, we consider the reconstructing to be inaccurate. We divide the number of accurate reconstructing over the entire number of repetitions to figure out the reconstructing percentage.  For all valid m and k values, the percentage of accurate data reconstructing for determined m and k in a 3D graph is presented and entered. The horizontal axis ratio of m/n is less than 1, its vertical axis ratio $k/m$ is also smaller than 1, and its height axis shows the accurate reconstructing for specific m and k. Furthermore, we present a precise definition of phase transition by giving two examples from phase transition graphs. 

In figure 5, a phase transition graph is presented for a matrix having a length $n=49$. AMP compression method has been used. The number of repetition for the purpose of obtaining the accurate reconstructing percentage is 200 times for every m and k. Moreover, the value of threshold to evaluate the accuracy of reconstructing is 50 db. In order to use AMP reconstructing method, the maximum considered input error is chosen to be 1e-8 and 200, to obtain an answer as well as the maximum number of repetition (in every reconstructing trial). 

\begin{figure}
\center
\label{AMP_PT}
\includegraphics[scale=0.75]{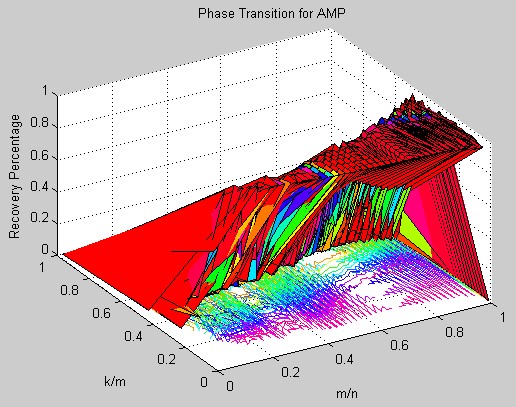}
\caption{Phase transition graph for AMP method and input data determined.}
\end{figure}

As it can be observed in figure 5, for every specific $m$, the AMP method executes the data reconstructing for signals $X_{n*1}$ by a sparse order having 100\% reconstructing. From a higher sparse order, the reconstructing order is reduced. If the number of repetition is chosen to be a large number to obtain reconstructing percentage, for every m, the reconstructing is done until a sparse order and accurate reconstructing percentage is 100\%. Furthermore, the reconstructing percentage suddenly becomes zero after that reconstructing order. The line that separates these two zones of 100\% reconstructing and zero percentage is called phase transition. This line is shown in figure 5.

In figure 6, the phase transition graph is presented for CoSamp reconstructing method. In this example, to find an answer for the main signal by CoSamp method, the matrix length is equal to 49, the number of repetition for obtaining the accurate reconstructing percentage is equal to 200. The value of threshold to evaluate the accuracy of reconstructing data is 50 db and the maximum error is considered to be 1e-8.

\begin{figure}
\center
\label{CoSamp_PT}
\includegraphics[scale=0.75]{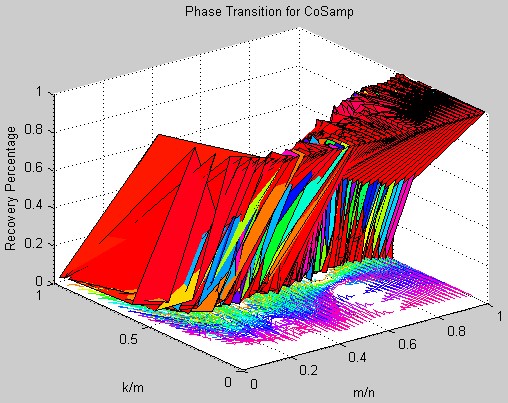}
\caption{Phase transition graph for CoSamp method and specified data input.}
\end{figure}

As it is observed, it seems that data reconstructing by CoSamp method is done for area greater than $m$ and $k$. In the next section, we discuss the other criterion to compare various reconstructing methods. 

\subsection{The ratio of signal to output noise to signal to input noise (SNRout-SNRin)}
\label{subsec42} 

In this section, we intend to draw one graph to compare the graphs for signal power. The ratio output noise to the input noise of signal power will be compared for various reconstructing methods. To draw this graph, we consider the dimensions of matrix $\Phi_{m*n}$ and also the order of sparseness of vector $X_{n*1}$ to be constant. In other words, numerical values of m, n, and k are constant. Assume that signal to input noise is a conventional number. In order to measure signal to output noise, it is necessary to generate sparse signal $X_{n*1}$ from K order for numerous repetitions. The multiplication resultant of $\Phi_{m*n} * X_{n*1}$ is equal to $Y_{m*1}$ which is calculated. We add this result with Guassian noise $(N_{m*1})$. To obtain the signal to input noise, it is necessary to calculate a coefficient before adding noise to compressed signal $Y_{m*1}$. The coefficient is obtained by the following equations after calculation of power of $Y_{m*1}$ and the power of noise.

\begin{equation}
\label{equ41}
\frac{P_s}{k^2*P_N} = SNR_{in} \ \ \ \Rightarrow \ \ \ k = \sqrt{\frac{P_s}{SNR_{in}*P_N}}.
\end{equation}

Therefore, the signal under reconstructing process is calculated by the following equation by the considered signal to noise. 

\begin{equation}
\label{equ42}
Y_{m*1_n} = Y_{m*1} + k*N_{m*1}.
\end{equation}

By using the reconstructing methods, the noise signal $Y_{m*1_n}$ will be reconstructed. In this case, signal $X_{n*1_n}$ is obtained. The reconstructed signal has noise and since we have signal $X_{n*1}$, we can obtain the signal to output noise. As it is stated earlier, the operation is would be repeated to obtain signal to output noise. In this case, by calculating the average of signal to output noises, a good average of the ratio of signal to output noise to signal to input noise can obtained. If continue this trend for various signal to input, we can draw the graph of ratio of signal to output noise to signal input noise for one or several reconstructing methods.

As a sample, the graph of ratio of signal to output noise to signal input noise is presented in figure 7 for two reconstructing methods of Yall1 and IHT. In the simulation, $n, m,$ and $k$ are selected to be equal to 49,25 and 10 respectively. In addition, the signal to input noise is considered to be between 40 db and 100 db. The signal variance and noise are generated stochastically and is equal to one. The number of repetition for calculating the ratio of signal to output noise to signal to input noise constant is 1000 times. The number of signal to input noise is chosen to be 30 by having an equal distance from each other within the range of 40 db and 100 db. The maximum error for finding the main signal answer in Yall1 method is 1e-7 with the maximum number of repetition 1000 (for obtaining an answer in every reconstructing trial).

\begin{figure}
\center
\label{IHT}
\includegraphics[scale=0.75]{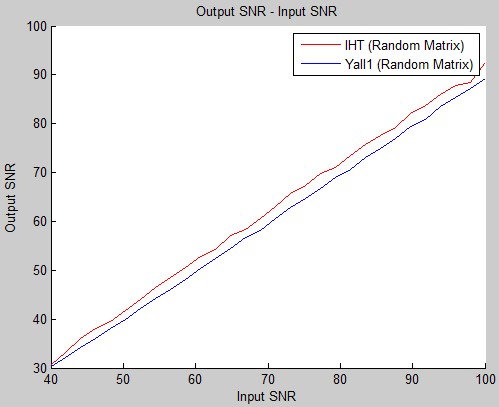}
\caption{The graph of ratio of signal to output noise to signal to input noise for both reconstructing methods of Yall1 and IHT.}
\end{figure}

As it is observed, by the input data given by two methods of IHT and Yall1, the signal to output noise is greater for IHT method compared to Yall1 method. Therefore, IHT method is more efficient compared to Yall1 method. In fact, the execution time of these two methods are not compared.  It is possible that the reconstructing timing is higher in IHT method compared to Yall1 method by the data input given to both methods. A more precise comparison can be done when the execution timing between these two methods are compared as well. If we consider input 1e-7 as the error of Yall1 and 1000 repetition as the input for method IHT, then the IHT method has higher efficiency.

In figure 8, the graph of ratio of signal to output noise to signal to input noise is presented for two methods of OMP\_enhanced and COSamp\_enhanced. To use these two methods, the matrix $\Phi_{m*n}$ is generated deterministically, because the reconstructing algorithm in OMP\_enhanced and CoSamp\_enhanced are designed and generated deterministically for matrices $\Phi_{m*n}$. In simulation, $n$ and $k$ are chosen to be 125 and 8. Since matrix is generated deterministically, m is determined for matrix $\Phi_{m*n}$ with respect to predetermined algorithm which generates $\Phi_{m*n}$. The numerical value of m is not available for the user in this case. To produce matrix stochastically, besides $n$, the numerical value of mu which is the maximum coherence between columns of matrix $\Phi_{m*n}$ should be determined. In addition, input $p$ which should be the first number representing the elements of $\Phi_{m*n}$ which are formed by how many separate components needs be determined as well. For example, if $p$ chooses the numerical value of 2, matrix $\Phi_{m*n}$ elements are formed from 1 and -1 or a coefficient of 1 and -1. Moreover, if $p$ chooses the numerical value of 3, the matrix $\Phi_{m*n}$ elements are formed by various elements which the numerical value of 3 is a distinct. In this simulation, the maximum coherence between columns of matrix $\Phi_{m*n}$ is equal to 1 and $p$ is designated to be 5. In this case, the number of matrix $\Phi_{m*n}$ rows is obtained to be 24. The signal to input noise interval is assigned to be 40 db and 100 db. The signal variance and noise are generated stochastically and is equal to 1 and the number of repetition for measuring signal to output noise for signal to constant input noise is 1000 times. The number of signal to input noise is equal to 30 which is chosen at an equal distance between 50 db and 100 db. The maximum error for finding the main signal answer is equal to 1e-8 in CoSamp\_enhance method. Matrix $\Phi_{m*n}$ is generated deterministically.

\begin{figure}
\center
\label{enhanced}
\includegraphics[scale=0.75]{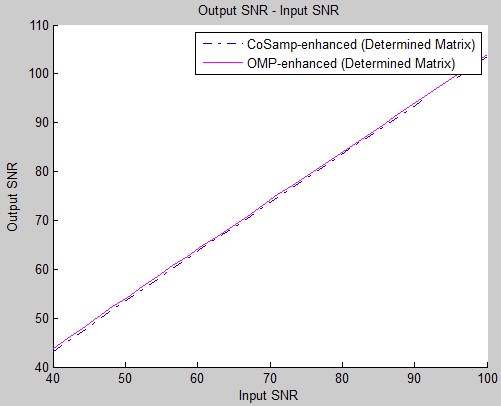}
\caption{The graph of ratio of signal to output noise to signal input noise for both methods OMP\_enhaced and CoSamp\_enhance.}
\end{figure}

As it is observed in figure 8, the signal to output noise for both methods of OMP\_enhanced and CoSamp\_enhanced does not differ considerably. In fact, it is observed that by using OMP\_enhanced method, signal to noise is a little more than when we use CoSamp\_enhanced method. However, this difference is not remarkable. Perhaps, it is too early to conclude that these two methods efficiencies do not differ greatly in compressed data reconstructing. We draw the graph of execution timing for every reconstructing method. The execution time that is needed to draw the above graph by using OMP\_enhanced, and CoSamp\_enhanced methods are 98 and 791 seconds. It is crystal clear that the signal to output noise is higher for OMP\_enhanced method compared to CoSamp\_enhanced method and also its execution time is greatly less than CoSamp\_enhanced method.

\subsection{The ratio of signal to noise in order of sparse signal $(k)$}
\label{subsec43}

The third criteria presented to compare the various reconstructing methods is the graph of ratio of signal to output noise to the order of sparseness of signal $(k)$. To draw this graph, we consider the matrix $\Phi_{m, n}$ dimensions to be constant. First, assume that the order of sparseness of signal $X_{n*1}$ is constant. $X_{n*1}$ signals by the order of sparseness are generated. First, we compress them by matrix $\Phi_{m, n}$, and then, as explained in previous section, (the equations \ref{equ41} and \ref{equ42}), we add a noise ratio to a specific signal to noise taken from a user to the signal $Y_{m*1}$ in order to obtain signal $Y_{m*1_n}$. Eventually, the signal $Y_{m*1_n}$ would be reconstructed by one of the reconstructing methods. Moreover, since the main signal $X_{n*1}$ is available, we can find signal to output noise. Estimation of the signal to output noise would be possible by calculating the average of the signal to output noises obtained in repetitions. If we change the order of signal sparseness $X_{n*1}$ from 1 to the maximum order of sparseness (which a user determines and intends to observe it on the graph), the graph of ratio of signal to output noise to sparseness order could be generated. We intend to draw the ratio of signal to noise to the order of sparseness for three methods. These three methods would be AMP, IHT and IMAT and then we deciphered the result. In order to do so, the dimensions of matrix $\Phi_{m*n}$ are considered to be as n=49, m=25, and the maximum sparseness order of signal $X_{n*1}$ is equal to 20. In addition, the number of repetitions to obtain the signal to output noise is assigned to be equal to 500. The signal to input noise is equal to 15 db, and signal variance which is generated stochastically, is equal to 1. Each one of the reconstructing methods have specific and separate inputs. In AMP reconstructing method, the maximum repetition to obtain an answer is equal to 200 and the maximum error to obtain an answer is assigned to be 1e-8. Regarding IHT reconstructing method, the maximum repetition to obtain an answer is designated to be 1000. In IMAT method, three inputs should be determined by the names of TO, iteration and alfa. The iteration input clearly shows the number of algorithm repetition which is equal to 10,000 in the simulation. However, the other two inputs of $T_0$ and $\alpha$ are determined with respect to input signal $X_{n*1}$ and its features. In the case which signals $X_{n*1}$ are generated stochastically by variance 1, these parameters are chosen to be 7.5 and 0.333 For the numerical values mentioned above, as the input parameters, the ratio of signal to output noise to the order of sparseness of signal $X_{n*1}$ is presented in figure 9. (Matrices are generated stochastically).

\begin{figure}
\center
\label{IMAT}
\includegraphics[scale=0.75]{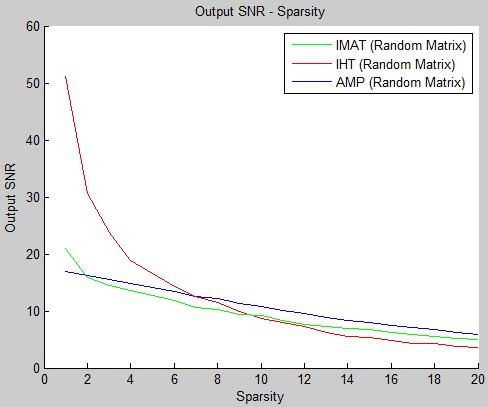}
\caption{The graph of ratio of signal to output noise to the order of sparseness of signal $X_{n*1}$ for reconstructing methods of AMP, IHT and IMAT.}
\end{figure}

With respect to figure 9, signal to output noise for the order of sparseness 1 is higher in the IHT reconstructing method in comparison to the other mentioned methods. Afterwards, IMAT and AMP methods respectively have more signal to output noises. In the interval between 2 and 6 for the sparse order, IHT, AMP and IMAT methods respectively allocate the most signal to noises to themselves. For the interval between 7 and 9 for the sparse order, the reconstructing methods AMP, IHT and IMAT respectively have the most signal to output noises. Finally, in high sparse orders between 10 and 20, AMP, IMAT and IHT respectively have more efficiency to obtain the signal to output noise.
In the next step, we examine OMP\_enhance and CoSamp\_enhanced methods. In this case, the numerical values of n, p and the maximum coherence of matrix columns $\Phi_{m*n}$ are assigned to be 125, 5, and 1. In this case, the numerical value of m for deterministic matrix $\Phi_{m*n}$ is equal to 24. The maximum order of sparseness of signal $X_{n*1}$ is chosen to be 20. The signal to input noise is selected to be 15 db and the number of repetition to obtain the signal to output noise is chosen to be 500. Moreover, the signal variance stochastically generated is chosen to be equal to 1. In addition, the specialized input of CoSamp\_enhanced method is assigned to be 1e-8 as well. In this case, the ratio of signal to noise to the order of sparseness of signal $X_{n*1}$ is illustrated in figure 9 for the both aforementioned methods. As it is observed in figure 10, in the small orders of both methods of OMP\_enhanced and CoSamp\_enhanced, they do not differ considerably from signal to noise point of view. However, by enlarging the order of sparseness, the OMP\_enhanced reconstructing method exhibits better efficiency. If we pay attention to the CoSamp\_enhanced reconstructing method, sudden and intense drops occur in the signal to output noise in sparse order 9.

\begin{figure}
\center
\label{OMP}
\includegraphics[scale=0.75]{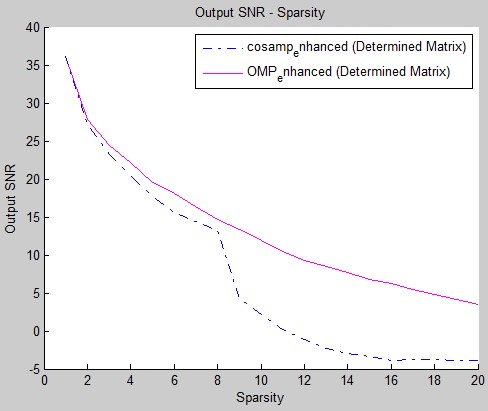}
\caption{The graph of ratio of signal to output noise to the order of sparseness of signal $X_{n*1}$ for OMP\_enhanced and COSamp\_enhanced reconstructing methods.}
\end{figure}

In the next section, we present the last criteria to compare various reconstructing methods.

\subsection{The ratio of percentage of reconstructing to the order of signal sparseness}
\label{subsec44}
The last criteria which we intend to examine is the ratio of reconstructing percentage to the order of sparseness of signal $X_{n*1}$. To draw this graph, the matrix dimensions ($m$ and $n$) are considered to be constant. We generate the signals repetition $X_{n*1}$ from $k$ order numerously.
Be careful that we presently consider the order of sparseness to be constant. We compress the signals $X_{n*1}$ by multiplying it by matrix $\Phi_{m*n}$. In this section, without adding compressed signal to noise, we reconstruct $Y_{m*1} = \Phi_{m*n} * X_{n*1}$ by one of the reconstructing methods. Since the main signal $X_{n*1}$ is available, we can obtain the signal to output noise because no noise is added to the compressed signal $Y_{m*1}$. We compare the signal to output noise with the numerical value of a threshold that the user has selected. If the signal to output noise is more than or equal to threshold, the reconstructing is precise and accurate. Otherwise, we consider the reconstructing to be inaccurate. One can obtain the percentage of reconstructing by calculating the ratio of number of accurate reconstructing to the number of all repetitions for the specific order of sparseness. If we change the order of sparseness of signal $X_{n*1}$ which is determined by a user, and also if we carry on the above procedures to obtain the reconstructing percentage for every order of sparseness, we can draw the graph of accurate reconstructing percentage based on the order of sparseness. Furthermore, we intend to draw and compare the graph of reconstructing percentage based on the order of sparseness for the reconstructing methods of AMP, OMP and CoSamp. For this task, the matrix $\Phi_{m*n}$ dimensions are chosen as n=49 and n=25. The maximum order of signal sparseness $X_{n*1}$ is equal to 20 and signal variance is assigned as 1 which is generated stochastically. In addition, the number of repetition is taken as 500 to obtain the reconstructing percentage for specific sparseness order. The special input of AMP method which is the maximum number of repetition as well as the maximum reconstructing algorithm error are chosen as 200 and 1e-8 respectively. Regarding reconstructing methods of OMP and CoSamp, the maximum reconstructing error is only defined as input which is equal to 1e-8. For all three reconstructing methods, the numerical value of threshold of signal to noise is assigned to be 50. The graph of the ratio of accurate reconstructing percentage to the order of sparseness of signal $X_{n*1}$ for input data given above is illustrated in figure 11.

By carefully examining the figure 11, it is observed that till the order of 4 of $X_{n*1}$ sparseness signal, no considerable difference is seen. As the order grows higher than 4, the reconstructing method of OMP has higher reconstructing percentage compared to the two other methods. From sparseness order 6 to 10. The CoSamp method yields better answer. However, between 11 to 14 the sparseness order changes and the reconstructing method of AMP functions better than CoSamp method. In fact, the reconstructing percentage is low for both methods. From the sparseness order higher than 15, none of the methods are able to reconstruct the signal $X_{n*1}$. (Matrix $\Phi_{m*n}$ are generated stochastically.)

\begin{figure}
\center
\label{COA}
\includegraphics[scale=0.75]{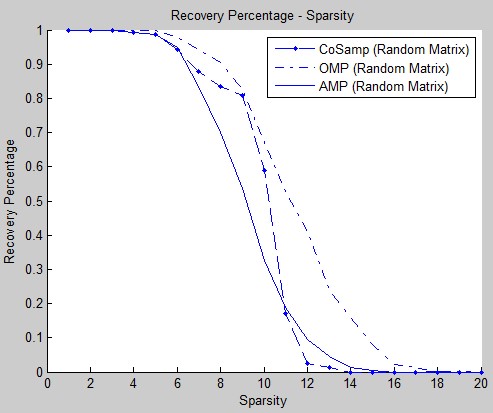}
\caption{The graph of ratio of accurate reconstructing percentages to the order of sparseness of signal $X_{n*1}$ for three methods of AMP, OMP and CoSamp.}
\end{figure}

In the next step, we draw and compare the graph of ratio of accurate reconstructing percentage to the order of sparseness of signal $X_{n*1}$ for two methods of OMP-enhanced and CoSamp\_enhanced. With respect to the fact that we should use deterministic matrices for both of these reconstructing methods, we allocate the numerical values of n, p and the maximum coherence of matrix $\Phi_{m*n}$ as 125, 5 and 1 respectively. After executing the algorithm to generate deterministic matrix, m is obtained as 24. The maximum order of sparseness of signal $X_{n*1}$ is chosen to be 20. The number of repetition to obtain the signal to output noise is chosen as 500. Moreover, signal variance is generated stochastically which is chosen to be equal to 1. Finally, the maximum error for reconstructing method of CoSamp is taken as 1e-8 (The OMP method has no need to have other input such as the number of algorithm repetition or the maximum reconstructing error). In figure 12, the graph of ratio of reconstructing percentage to the order of sparseness of signal $X_{n*1}$ for the above inputs are illustrated. (Matrix QM8N is generated stochastically)

\begin{figure}
\center
\label{OMP_e}
\includegraphics[scale=0.75]{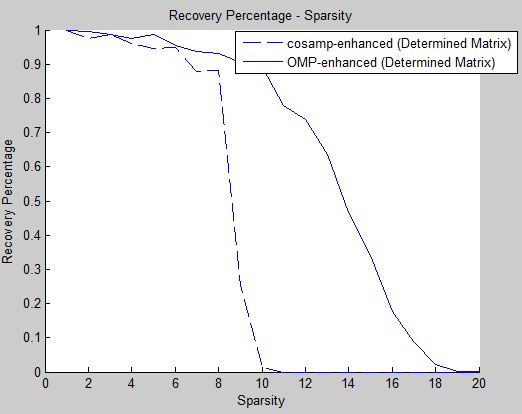}
\caption{The graph of ratio of accurate reconstructing percentage to the order of sparseness of signal $X_{n*1}$ for two methods of OMP\_enhanced and CoSamp\_enhanced.}
\end{figure}

It is crystal clear from figure 4-8 that the accurate reconstructing percentage for OMP\_enhanced is considerably better than Cosamp\_enhanced method. It is worth mentioning that in reconstructing method Cosamp\_enhanced, the reconstructing percentage for orders of sparseness higher than 9 intensely is dropped. This behavior was observed in drawing the graph of ratio of signal to output noise to the order of sparseness as well. As it was stated in previous section, not only the reconstructing method of OMP-enhanced functions better in reconstructing the compressed signal, but also the execution time is much better. Execution time of drawing the graph of ratio of reconstructing percentage to the order of sparseness for reconstructing method of OMP\_enhanced is nearly 62 seconds and this number for Cosamp\_enhancec method is equal to 492 seconds.

\section{The GUI Guide to Draw the Aforementioned Graphs in Chapter \ref{sec4}}
\label{sec5}

The ultimate goal of this project is to use the Graphical User Interface in MATLAB environment to draw the graphs discussed in chapter 4. The main goal is to compare these reconstructing methods easier from different aspects. Therefore, in this project, to facilitate the usage of MATLAB files,  a MATLAB environment in realm of GUI MATLAB is provided and shown in figure 13 \cite{Wiki7}.

\begin{figure}
\center
\label{GUI}
\includegraphics[scale=0.37]{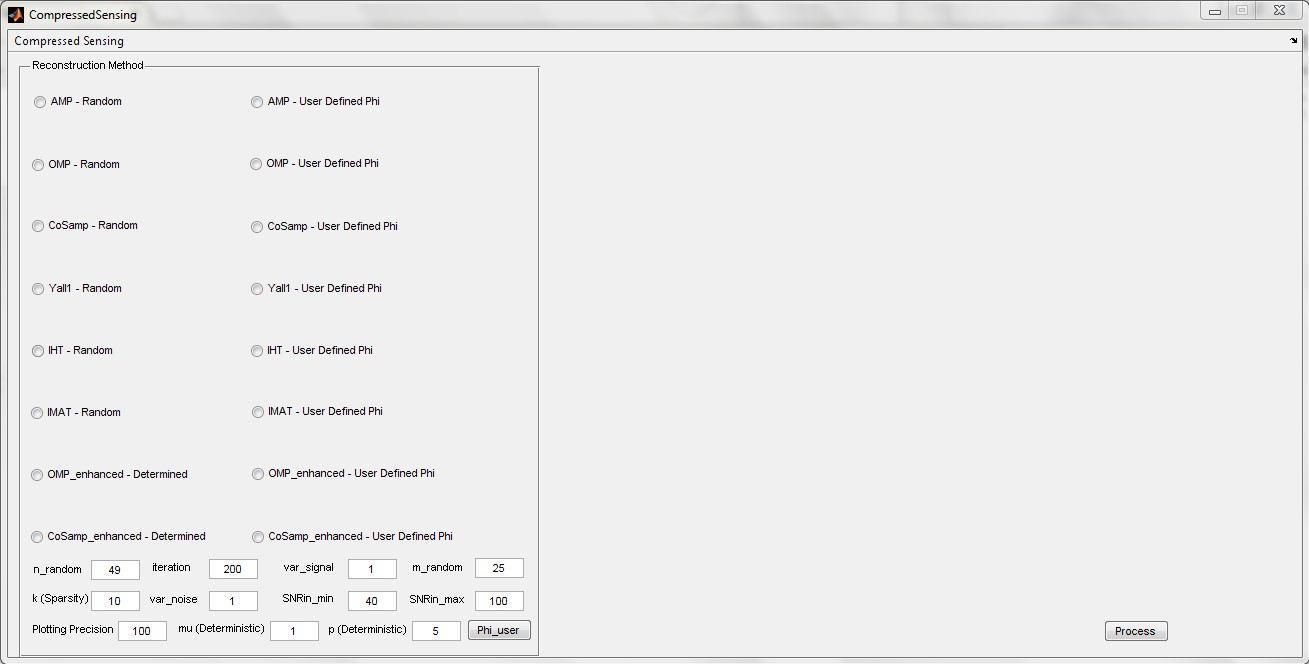}
\caption{General image of the designed GUI.}
\end{figure}

In the left menu and above GUI designed, the desirable graph among the graphs presented in chapter 4 could be selected. After selecting the desirable graph, various reconstructing methods are exhibited which can be selected. By choosing the reconstructing method from the first column, the matrix $\Phi_{m*n}$ can be generated stochastically or deterministically. However, by selecting reconstructing methods in the second column, the matrix $\Phi_{m*n}$ has to be updated. In the bottom and left side of the GUI, there are a series of common inputs which are all identical for all the reconstructing methods. However, by choosing any of reconstructing methods, a window would be open at the right side of the screen which receives the special inputs of reconstructing method. By selecting all the reconstructing methods in the first column, the GUI page is formed as figure 14.

\begin{figure}
\center
\label{GUI2}
\includegraphics[scale=0.37]{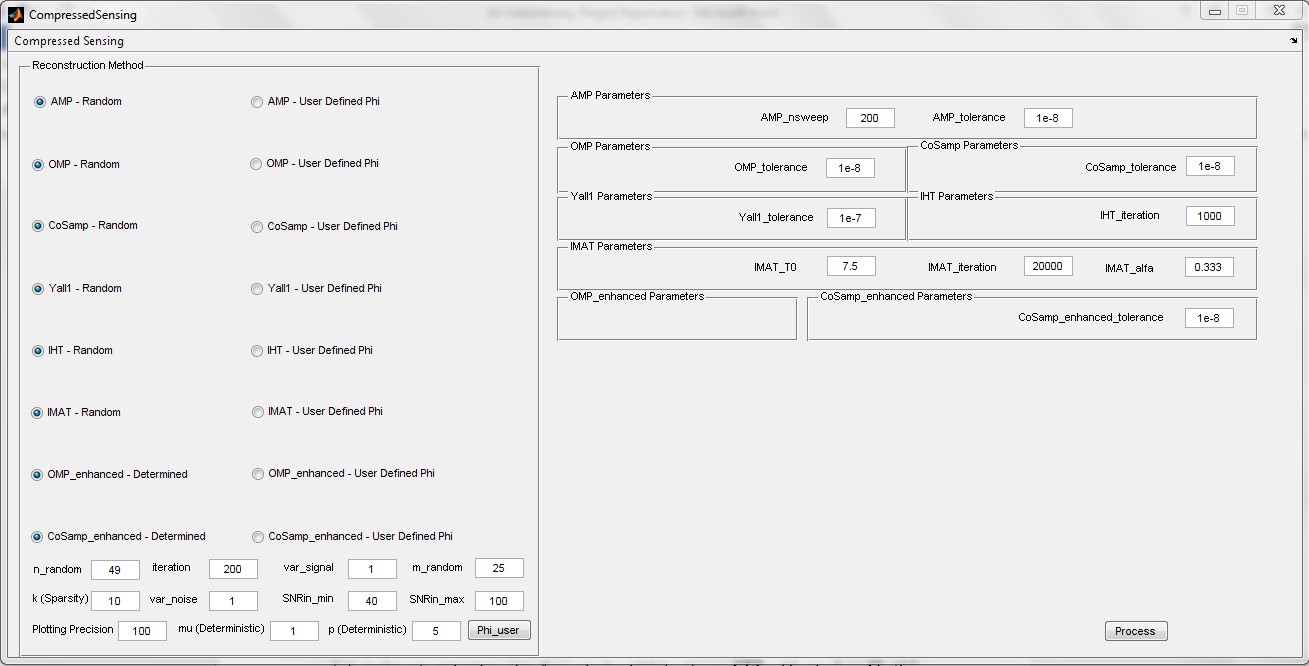}
\caption{The general image of GUI after choosing all the reconstructing methods in the first column.}
\end{figure}

Although the majority of inputs were explained along the report, as a comprehensive explanation and easier reference, we explain the entire inputs. The first input named as n\_random shows the number of columns of matrix $\Phi_{m*n}$ (for both stochastic and deterministic matrices). The m\_random input shows the number of columns of matrix $\Phi_{m*n}$ in the case in which matrix $\Phi_{m*n}$ is generated stochastically. The var\_signal input and var\_noise input show the signal variance $X_{n*1}$ and the signal to noise respectively. The input $k$ shows the order of sparseness of signals $X_{n*1}$, SNRin\_min, and SNR\_max. The mu shows the maximum coherence between matrix $\Phi_{m*n}$ columns for the scenario in which it is generated deterministically. The number p is also the first number showing that the matrix $\Phi_{m*n}$ elements can be assigned to several separate numerical values. If $P$ is equal to 2, the matrix $\Phi_{m*n}$ elements are formed from 1 to -1 or from one of their coefficients. Moreover, if P is equal to 3, the matrix $\Phi_{m*n}$ elements are formed by three mixed and separate numbers. The plotting precision input is applicable to draw the graph of ratio of signal to output noise to the signal to input noise. The signal to output noise is calculated for several various numerical values from the signal to input noise. To draw the graph of ratio of the signal to output noise to the order of sparseness of signal $X_{n*1}$, we have to determine the size of signal to input noise. This number is determined based on db in SNR-in input. To draw the two graphs of ratio of the signal to output noise to the order of sparseness and the ratio of reconstructing percentage to the order of sparseness, the order of sparseness of signal $X_{n*1}$ changes from 1 to its maximum numerical value. The aforementioned maximum numerical value is determined in the input called Sparsity\_max. The iteration input shows that how many iterations would occur in order to compute the signal to output noise or the reconstructing percentage. Finally, there is a button called Phi\_user as input which provides the possibility to upload desirable matrix $\Phi_{m*n}$ and also to draw the graph for the matrix. After choosing any of the reconstructing methods, a window is exhibited at the right side of the screen which shows the reconstructing method. For reconstructing methods of AMP, IHT and IMAT, we have the inputs of AMP-nsweep, IHT-iteration and IMAR-iteration which exhibit the maximum repetition of AMP, IHT and IMAT algorithms. In addition, the reconstructing methods of AMP, OMP, Yall1 and CoSamp\_enhanced have the inputs of AMP-tolerance, OMP\_tolerance, CoSamp\_tplerance, Yall1\_tolerance and CoSamp\_enhanced\_tolerance which exhibit the maximum error of the reconstructed signals. The reconstructing method of OMP\_enhanced has no special input. However, the IMAT method has two more inputs called IMAT\_T0 and IMAT\_Alfa. IMAT\_T0 and IMAT\_Alfa differ for various signals of $X_{n*1}$ with different features. Their numerical values of these inputs have to be adjusted manually. If we generate the signals $X_{n*1}$ stochastically by variance 1, these two inputs are selected to be equal to 7.5 and 0.333 respectively. By choosing the specific graph from the above menu and selecting the reconstructing methods with specific inputs, the results would be displayed after choosing the ``Process'' bottom at the right side and bottom of the screen.

%\appendix

%\section{Section in Appendix}

%\label{appendix-sec1}

%% References

\section{Conclusion and Future Work}
In this work, we studied different deterministic compressed sensing approaches including AMP, IHT, IMAT, YALL1, CoSamp, and compared them against each other and developed a GUI interface for the user to be able to compare these algorithms. Sparse signals have extensive applications in physiological signal processing \cite{counts2007were, lederer2013distributed, riiser2013commute, hosseini2017mobile, alrefaie2014road, chehardeh2018closed}, communication \cite{haddadpour2017simulation}, text compression \cite{kavousi2018estimating}, image processing \cite{yao2018multifractal, yao2018markov, chehardeh2016remote}, and etc., where the mentioned compressing methods in this email can be used for a better compression. For future work, one can focus on the different applications of sparse signals and apply the deterministic compressed sensing to them.

\vspace{1cm}

\bibliographystyle{elsarticle-num}

\bibliography{sample}

\end{document}